\documentclass[twocolumn,superscriptaddress,nofootinbib,aps,prl,floatfix,preprintnumbers,amsmath,amssymb,groupedaddress]{revtex4}

\usepackage{dsfont}
\usepackage{epsfig}
\usepackage{slashed}
\usepackage{bbold}
\usepackage{psfrag}
\usepackage{color}
\PassOptionsToPackage{caption=false}{subfig}
\usepackage{subfig}
\usepackage{multirow}
\usepackage{booktabs}
\usepackage{hyperref}
\usepackage{pstricks}
\usepackage{feynmp}

\newcommand{\beq}{\begin{equation}}
\newcommand{\eeq}{\end{equation}}
\newcommand{\bea}{\begin{eqnarray}}
\newcommand{\eea}{\end{eqnarray}}

\renewcommand{\O}{\mathcal{O}}

\def\lpar#1#2#3#4{\rlap{\raise#3\hbox{$\hskip#4#1\left\{\mbox{\phantom{\rule[0mm]{0mm}{#2}}}\right.$}}}
\def\rpar#1#2#3#4{\rlap{\raise#3\hbox{$\hskip#4\left\}#1\mbox{\phantom{\rule[0mm]{0mm}{#2}}}\right.$}}}

\begin{document}
 \title{Dynamical R-Parity Violation}

\author{Csaba Cs\'aki}
\affiliation{Department of Physics, LEPP, Cornell University, Ithaca, NY 14853, USA}
\author{Eric Kuflik}
\affiliation{Raymond and Beverly Sackler School of Physics and Astronomy, Tel-Aviv University, Tel-Aviv 69978, Israel}
\author{Tomer Volansky}
\affiliation{Raymond and Beverly Sackler School of Physics and Astronomy, Tel-Aviv University, Tel-Aviv 69978, Israel}

%\pacs{PACS}
%\keywords{keywords}
%\preprint{UCB-PTH-13/?? }
\begin{abstract}
We present a new paradigm for supersymmetric theories with R-parity violation (RPV).   At high scale, R-parity is conserved in the visible  sector but spontaneously broken in the SUSY-breaking sector.  The breaking is then dynamically mediated to the visible sector and is manifested via non-renormalizable operators at low energy.  Consequently, RPV operators originate from the K\"ahler potential rather than the superpotential, and are naturally suppressed by the SUSY-breaking scale, explaining  their small magnitudes.   A new set of non-holomorphic RPV operators are identified  and found to often 
dominate over the standard RPV ones.  We study   the relevant low-energy constraints arising from baryon-number violating processes, proton decay and flavor changing neutral currents, which may all be satisfied if a solution to the Standard Model flavor puzzle is incorporated. The chiral structure of the RPV operators imply new and distinct collider signatures, indicating the need to alter current techniques in searching for RPV at the LHC.
\end{abstract}
\maketitle
%%%%%%%%%%%%%%%%%%%%%%%%%%%%%%%%%%%%%%%%%%%%%%

%%%%%%%%%%%%%%%%%%%%%%%%%%%
\section{Introduction} 
%%%%%%%%%%%%%%%%%%%%%%%%%%%

Supersymmetry (SUSY) has long been considered to be the leading candidate for solving the hierarchy problem.  However, searches in the first three years of the LHC  have failed to uncover evidence for the existence of superpartners, thereby severely constraining the parameter space of the Minimal Supersymmetric Standard Model (MSSM) and pushing the masses of some of the superpartners to uncomfortably high scales.  Thus, if supersymmetry is to remain natural, it must manifest itself differently than in standard scenarios.

 The vast majority of SUSY searches  study events with significant missing energy, as typically follows from the implicit assumption of R-parity conservation.   A way to evade many of the  bounds is to consider theories in which R-parity is violated~\cite{RPV}. Traditionally RPV models introduce the following holomorphic operators:
\begin{eqnarray}
{\cal  O}_{\rm hRPV}&=& \frac{1}{2}\lambda_{ijk}  L_iL_j\bar{e}_k +\lambda^\prime_{ijk}  L_i Q_j \bar{d}_k+ \frac{1}{2}\lambda^{\prime\prime}_{ijk} \bar{u}_i\bar{d}_j\bar{d}_k \,, \nonumber \\
{\cal O}_{\rm hBL}&=& \mu_i L_i H_u \,,
\label{eq:WRPV}
\end{eqnarray}
where ${\cal O}_{\rm hRPV}$ are the trilinear terms that do not contain dimensionful parameters, while ${\cal O}_{\rm hBL}$ are the holomorphic  bilinear RPV terms, with dimensionful couplings $\mu_i$. 
These operators are usually written in a superpotential, $W_{\rm RPV} ={\cal O}_{\rm hRPV}+{\cal O}_{\rm hBL}$. 
The above couplings, however, are strongly constrained as they generically allow for rapid proton decay, di-nucleon decays, neutron-anti-neutron oscillations, flavor changing processes and cosmological depletion of any baryon asymmetries (for a review, see~\cite{RPVreview}).    Thus RPV theories must incorporate extremely small and seemingly {\it ad hoc} couplings.  

 Recently, a proposal for an organizing principle that could explain the smallness and hierarchical nature of the RPV couplings  above  was introduced~\cite{MFVSUSY} (see also~\cite{Smith,Dreiner}), based on the Minimal Flavor Violation (MFV) principle, whereby the magnitude of the RPV couplings is related to the small Yukawa couplings of the flavor sector,   naturally generating a hierarchy which leads to a viable pattern of RPV.  Related models as well as recent studies on the LHC phenomenology of baryonic RPV models can be found in~\cite{moreRPVflavor,  Krnjaic:2013eta,RPVLHC,Ruderman:2012jd}. 
 
 The main goal of this paper is to present an alternative to the traditional approach summarized in~\eqref{eq:WRPV}, by postulating a dynamical origin of RPV.    In particular, the visible sector is assumed to be R-parity conserving, while its breaking, which occurs in a hidden sector, is dynamically communicated to the visible sector.
   An immediate consequence is that RPV-inducing operators {\em naturally appear in the K\"ahler potential}, and are  suppressed by the mediation scale, while they may or may not appear in the superpotential.   As a result, under some quite general and natural circumstances, the terms
  in Eq.~\eqref{eq:WRPV} are 
  not the leading set of RPV operators and are
  insufficient to describe the low-energy dynamics of the model.   In particular, {\em  new types of RPV operators with distinct phenomenology naturally arise} and must be considered in any search for RPV supersymmetry.
 
While not necessarily related, it is interesting to postulate a joint mechanism  for breaking and mediating both supersymmetry and R-parity   (for related ideas see~\cite{Krnjaic:2013eta}). 
Since R-parity in the visible sector is equivalent to $(-1)^{3(B-L)+2s}$, where $s$ is the spin of the particle, the sector which triggers the breaking must be charged under that symmetry too.  It is then also natural to consider a flavor-dependent mediation mechanism, such as the  so-called flavor mediation models based on the Froggatt-Nielsen (FN) mechanism~\cite{FN,flavormediation}, or those which allude to partial compositeness.    In such scenarios, additional suppression of the RPV terms is obtained, along the lines mentioned above. These suppressions will typically  be present even if the flavor model is unrelated to the mediation scheme.

In what follows we make the following assumptions:
\begin{enumerate}
\item[I.] {\bf Dynamical RPV (dRPV)}.  RPV is broken dynamically in a hidden sector. 
\item[II.] {\bf RPV is related to SUSY breaking}.
\end{enumerate}
These assumptions then imply the appearance of novel non-holomorphic RPV operators, 
\begin{eqnarray}
\label{eq:KRPV}
{\cal O}_{\rm nhRPV} &=&  \eta_{ijk} \bar{u}_i\bar{e}_j\bar{d}_k^\dagger + \eta^\prime_{ijk} Q_i\bar{u}_j L_k^\dagger + \frac{1}{2}\eta^{\prime\prime}_{ijk} Q_i Q_j \bar{d}_k^\dagger 
\nonumber \\ &&+\ \kappa_i \bar{e}_i H_d H_u^\dagger \,,
\\
{\cal O}_{\rm nhBL} &=& \kappa'_i L_i^\dagger H_d\, ,
\label{eq:KRPV2} \end{eqnarray}
 which can show up in the K\"ahler potential, coupled to a  SUSY-breaking spurion $X=M + \theta^2 F_X$. Here we define all the couplings 
  to be dimensionless. 
 The main consequence of these assumptions is that all RPV interactions will automatically be suppressed by, at least,  
\begin{equation}
\epsilon_X\equiv F_X/M^2\,,
\end{equation}
 which may vary in size from ${\cal O}(1)$ to ${\cal O}(10^{-16})$ as in gravity mediation.   Its smallness may explain why all of these terms are very small to start with.

We may further assume:   
\begin{enumerate}
\item[III.] {\bf Dynamical solution to the SM flavor hierarchy.}
\end{enumerate}
 With this third assumption additional, flavor-dependent suppression factors arise.  One then obtains a natural organizing principle which generates  a  hierarchy in the RPV couplings.   Indeed any solution to the flavor hierarchy, such as the above mentioned FN model or partial compositeness, can be incorporated and would typically produce similar hierarchy in the RPV operators.

The RPV operators related to (\ref{eq:KRPV}) and (\ref{eq:KRPV2}) have not been studied before. We will argue below that the above three assumptions are sufficient to suppress any flavor-violating transitions, and in particular proton decay, without assuming lepton-number conservation.    Moreover, the new  operators predict novel and distinct LHC signatures.  In this paper we study the basic constraints and phenomenology of the above new operators, demonstrating their unique features, as well as the viability of this scheme.  A more detailed LHC study and a UV complete model of dRPV will appear in upcoming publications~\cite{future1}.

%%%%%%%%%%%%%%%%%%%%%%%%%%%%%%%%%%%%%%
\section{Framework}
%%%%%%%%%%%%%%%%%%%%%%%%%%%%%%%%%%%%%%

In accordance with assumptions (I) and (II) discussed above, we consider a two-scale scenario in which the single spurion, $X$, 
breaks both R-parity and supersymmetry in a hidden sector, while providing the messenger mass scale.   We will see below that $F_X/M^2\ll~1$ is preferable, following constraints on RPV operators.   We further assume that $M \ll  M_{\rm Pl}$ for the mediator scale.

To derive constraints on dRPV, its low energy description must be understood.  As is customary when studying supersymmetry breaking, the low energy  SM Lagrangian is assumed to be  accompanied by the above spurion $X$, parametrizing the effects of the hidden sector.   Depending on the UV completion, $X$ may be charged under various continuous and discrete symmetries which will constrain its  low-energy effective couplings.   Nonetheless, a low-energy analysis suffices to restrict the form of the RPV operators which may show up.  Indeed, one may  assume that  $B-L$ is  preserved at low energy in the visible sector, as is typically the case.
 In order to break R-parity, $X$ must  then be charged under $B-L$, while we will also consider the possibility that it is additionally charged under an unbroken  $U(1)_R$ symmetry. 
  
As a consequence, since $\O_{\rm nhRPV}+\O_{\rm nhBL}$ and $\O_{\rm hRPV}+\O_{\rm hBL}$ are  charged $+1$ and $-1$  under $B-L$, respectively, they are distinguishable at  low energy.  If, for example, $X$ is charged $-1$ under $B-L$, the K\"ahler potential and superpotential take the following form at leading order:
\begin{eqnarray}
K_{\rm dRPV} &=& \frac{1}{X^\dagger} \O_{\rm nhRPV} + \frac{X}{M_{\rm Pl}}\O_{\rm nhBL}  \nonumber \\ &&+  \frac{X^\dagger}{M_{\rm Pl}^2} \left(\O_{\rm hRPV} + \O_{\rm hBL}\right) +h.c. \,,
\label{eq:1}
\\ 
W_{\rm dRPV} &=& \frac{X}{M_{\rm Pl}^2} \left( \rho_{ijk} H_d Q_iQ_jQ_k + \rho'_{ijk} H_d Q_i \bar{u}_j \bar{e}_k \right)
\label{eq:1b}
\end{eqnarray}
   Note that the K\"ahler term $\frac{1}{X} \O_{\rm hRPV}$ is removed by a K\"ahler transformation, while the term $X \O_{\rm nhRPV}/M_{Pl}^2$ is subleading.  We thus find that in this case the holomorphic RPV operators, when generated dynamically, are highly suppressed  in comparison to the new non-holomorphic cubic ones.   Furthermore, the non-holomorphic bilinear terms are also suppressed and their effect is negligible as discussed below.

\begin{figure*}[t!]
\begin{center}
\begin{fmffile}{nnbar}
	      \begin{fmfgraph*}(170,60)
	     \fmfstraight
	        \fmfleft{i3,i2,i1}
	        \fmfright{o3,o2,o1}	        
	        \fmf{fermion}{i1,v1}
	        \fmf{fermion}{i2,v1}
                 \fmf{fermion}{v3,i3}
                 \fmf{fermion,tension=1,label=$\tilde{g}$,label.side=left}{v7,v3}
  \fmf{fermion,tension=1,label=$\tilde{g}$}{v7,v6}
                 \fmf{wiggly,tension=0}{v3,v6}
                 \fmf{fermion}{v6,o3}
                 \fmf{fermion}{o1,v4}
                 \fmf{fermion}{o2,v4}
                 \fmf{phantom}{v1,v4}

                 \fmffreeze

                 \fmf{scalar,label=$\tilde{d}_R$}{v1,v3}

                 \fmf{scalar,label=$\tilde{d}_R$}{v4,v6}
\fmffreeze
%\fmf{gluon}{v3,v6}
                 
                  \fmflabel{$Q_{u}$}{i1}
	        \fmflabel{$Q_{d}$}{i2}
	        \fmflabel{$\bar{d}^\dagger$}{i3}
	        \fmflabel{$Q_{u}$}{o1}
	        \fmflabel{$Q_{d}$}{o2}
	        \fmflabel{$\bar{d}^\dagger$}{o3}
\fmfv{decoration.shape=cross,decoration.angle=0,decoration.size=3thick}{v7}
%	                      \fmfv{decoration.shape=cross,decoration.angle=0,decoration.size=3thick}{v5}
	        \end{fmfgraph*}
	    \end{fmffile}  
	    \hspace{3cm}
	 \begin{fmffile}{FCNC1}
	        \begin{fmfgraph*}(70,50)
	     \fmfstraight
	        \fmfleft{i1,i2}
	        \fmfright{o1,o2}
                 \fmf{fermion}{i2,v1}
                 \fmf{fermion}{i1,v1}
                 \fmf{scalar,label=$\tilde{d}_k$}{v1,v2}
                 \fmf{fermion}{v2,o2}
                 \fmf{fermion}{v2,o1}

                  \fmflabel{$Q_i^\beta$}{i2}
	        \fmflabel{$Q_i^\alpha$}{i1}
	        \fmflabel{$Q_j^{\alpha \dagger}$}{o2}
	        \fmflabel{$Q_j^{\beta \dagger}$}{o1}
 
	        \end{fmfgraph*}
	    \end{fmffile}  \hspace*{0.5cm} \begin{fmffile}{FCNC2}\hspace*{9mm}
	        \begin{fmfgraph*}(70,50)
	     \fmfstraight
	        \fmfleft{i1,i2}
	        \fmfright{o1,o2}
                 \fmf{fermion}{i2,v1}
                 \fmf{fermion}{i1,v1}
                 \fmf{scalar,label=$\tilde{L}_k$}{v1,v2}
                 \fmf{fermion}{v2,o2}
                 \fmf{fermion}{v2,o1}

                  \fmflabel{$\bar{u}_j^\alpha$}{i2}
	        \fmflabel{$Q_i^\alpha$}{i1}
	        \fmflabel{$u_i^{\beta \dagger}$}{o2}
	        \fmflabel{$Q_j^{\beta \dagger}$}{o1}
 
	        \end{fmfgraph*}
	    \end{fmffile}  	
	    \end{center}
\caption{\label{deltab2} {\bf Left}: An example diagram for $\Delta B =2 $ processes. The diagram induces $n-\bar{n}$ oscillations and dinucleon decay. {\bf Right}: RPV contributions to the $\Delta F =2 $ operators $\mathcal{Q}_1^{q_i q_j} \equiv -\frac{1}{2}  (Q_i^\alpha Q_i^\beta) (Q_j^{\alpha\dagger} Q_j^{\beta\dagger})$ [left] and $\mathcal{Q}_4^{q_i q_j} \equiv \bar{u}_j^\alpha Q_i^\alpha Q_j^{\beta\dagger} \bar{u}_i^{\beta\dagger}$ [right].}
\end{figure*}

If instead $X$ has charge $+1$ under $B-L$, then the leading holomorphic RPV operator is $\frac{1}{X^\dagger} \O_{\rm hRPV}$, while the leading non-holomorphic term is $\frac{1}{X} \O_{\rm nhRPV}$.
At this stage the two terms appear to be of the same order, however the non-holomorphic terms might still be suppressed 
due to their chiral structure.  For instance,  the $QQ\bar{d}^\dagger$ operator  will induce couplings which are suppressed by $m_d/M $,  compared with the $F_X/M^2$ suppression of $\O_{\rm hRPV}$.   Similar conclusions are obtained for other choices of charges under $B-L$. 

If $X$ is further charged under a $U(1)_R$ symmetry (assuming flavor-independent charges), it is impossible to generate all of the holomorphic and non-holomorphic RPV operators at the same time. Thus, depending on the $X$ R-charge, only some of the above 
will appear at low energy.    
Finally, if  the $B-L$ symmetry is instead promoted to an R-symmetry, one finds once again that the non-holomorphic RPV operators dominate. 

We therefore conclude that in the absence of additional scales, dRPV allows for either the holomorphic or the non-holomorphic RPV operators to be generated,  
but {\em the non-holomorphic ones should not be neglected}.  Given that previous studies consider  exclusively holomorphic RPV, we will study below the case when only the non-holomorphic RPV terms appear.

Before analyzing the constraints, let us briefly discuss assumption (III).     The inclusion of flavor dynamics implies that the various operators discussed above are suppressed according to their flavor structure.   Numerous models that  introduce such suppressions exist, including, for example, theories with horizontal symmetries    as in FN  models~\cite{FN}, or ones with strong interactions~\cite{RS, NelsonStrassler, Rattazzi}.    
Consequently, the  low energy parameters, $\eta$, $\eta^\prime$, $\eta^{\prime\prime}$, $\kappa$ and $\kappa^{\prime}$, are suppressed in a flavor-dependent manner.  For example, the $\eta^{\prime\prime}_{ijk}$'s  can take the form 
\begin{equation} \label{eq:3}
\eta_{ijk}^{\prime\prime} \sim \epsilon^{|q_{Q_i}+q_{Q_j}-q_{d_k}|}\,,
\end{equation}
where $\epsilon={\cal O}(0.1)$ is a small parameter and $q_\alpha$ are the various charges of the SM fields under the FN symmetry.  Similar expressions hold when $q_\alpha$ characterize the partial compositeness in the case of an RS-type scenario.
While a comprehensive study is beyond the scope of this paper, we stress that all the constraints discussed below are easily satisfied with, for example, a simple choice of FN charges.   In particular, a straightforward extension of the alignment model of~\cite{NirSeiberg} to the lepton sector allows for a viable dRPV model, without any additional assumption such as  the typically needed lepton-number conservation.    A  complete realization of this scenario will be discussed  in an upcoming publication~\cite{future1}.

Finally a remark is in order.   Assumption (III) may require introducing an additional spurion (such as the one responsible for breaking the FN symmetry).   
A spurion of this kind  may modify the  above discussion which is based on the existence of just two scales, $X$ and $M_{\rm Pl}$, and as a result the suppression of the holomorphic RPV operators may naively be milder.    Complete models, however, will typically include additional symmetries which can  
forbid the holomorphic operators altogether~\cite{future1}.  

%%%%%%%%%%%%%%%%%%%%%%%%%%%%%%%%%%%%%%
\section{Low energy constraints}
%%%%%%%%%%%%%%%%%%%%%%%%%%%%%%%%%%%%%%
The operators in~\eqref{eq:KRPV} violate baryon (B)- and/or lepton-number (L), in addition to the non-abelian SU(3)$^5$ flavor symmetries of the SM.   As a result, low energy bounds exist, which we derive below.   As mentioned above, all these bounds are easily satisfied with the inclusion of a simple flavor model.

%%%%%%%%%%%%%%%%%%%%%%%%
\subsection{$\Delta B =2 $ Processes}
%%%%%%%%%%%%%%%%%%%%%%%%
The $\eta''$ term in~\eqref{eq:KRPV} violates baryon-number (B) by one unit.  Consequently it is important to check that the bounds on $\Delta B=2$ processes, $n-\bar{n}$ oscillations and dinucleon decay, obtained by two  insertions of this  vertex, are obeyed. The simplest way is to integrate out the squarks which will generate a dimension-9 operator.   While the most general flavor index structure is allowed, we here display the subset necessary for the constraints.  Considering the left diagram of Fig.~\ref{deltab2} one finds, 
\beq
\frac{1}{\Lambda^5_{ijk}} (Q_i Q_i Q_j Q_j \bar{d}_k^\dagger \bar{d}_k^\dagger)\,,
\label{dim9op}
\eeq
with the suppression scale,
\beq
\dfrac{1}{\Lambda_{ijk}^{5}} = \pi \alpha_s \dfrac{\eta^{\prime\prime}_{iik} \eta^{\prime\prime}_{jjk}}{m_{\tilde{g}} m^4_{\tilde{d}._{R,k}}} \epsilon_X^2\,. 
\eeq
This leads to $n-\bar{n}$ oscillations and dinucleon decay $pp\rightarrow \pi^+ \pi^+$ for $i,j,k=1$ and $pp\rightarrow K^+ K^+$ for ${i,j=1;k=2}$. 

The $n-\bar{n}$ oscillation time is approximately given by 
\beq
\tau_{n-\bar{n}} \simeq \frac{\Lambda_{111}^5}{2\pi \tilde{\Lambda}^6_{QCD}}\,,
\eeq
where $ \tilde{\Lambda}_{QCD}$ is the hadronic matrix element which we 
 estimate  at $200$ MeV. We find,
\beq
\tau_{n-\bar{n}} \simeq 3 \times 10^8 {\rm ~s}   \left(\frac{m_{\tilde{d}_{R1}}}{{\rm TeV}} \right)^4  \left(\frac{m_{\tilde{g}}}{{\rm TeV}} \right)
\left(\frac{4 \times 10^{-2}}{\eta^{\prime\prime}_{111}} \right)^2 \left( \frac{10^{-5}}{\epsilon_X}\right)^2\,,
\eeq 
to be compared with the experimental bound ${\tau_{n-\bar{n}} > 2.44\times10^8}$ s \cite{Abe:2011ky}.

The same operator also contributes to the dinucleon decay process $pp\rightarrow \pi^+\pi^+ (K^+K^+)$. The approximate expression for the width is given by~\cite{Sher},
\beq
\Gamma \simeq  \frac{8}{\pi} \frac{\rho_N}{m_N^2}\frac{\tilde{\Lambda}^{10}_{QCD}}{\Lambda^{10}_{pp}}\,,
\eeq
where $\rho_N \simeq0.25 {\rm~fm}^{-3}$ is the nuclear matter density and $\Lambda_{pp} \equiv \min\{\Lambda_{11k},\Lambda_{1k1}\}$ under the assumption that only one operator dominates the process.   Here $k=1$ or 2, depending on whether the decay is to pions or kaons. The bound on the lifetime is $\tau_{pp} \geq1.7\times 10^{32}$ years \cite{Litos:2010zra} while in our model we find,
\beq
\tau_{pp} \simeq 5 \times 10^{32} {\rm ~yr}  \left(\frac{m^8_{\tilde{d}_{R,k}}m^2_{\tilde{g}} }{{\rm TeV}^{10}} \right) \left(\frac{10^{-1}} {\eta^{\prime\prime}_{pp}} \right)^4 \left( \frac{10^{-5}}{\epsilon_X}\right)^4 \,.
\eeq  
where $\eta^{\prime\prime}_{pp}\equiv{\max\{\eta^{\prime\prime}_{11k},\eta^{\prime\prime}_{1k1}\}}$. 

%%%%%%%%%%%%%%%%%%%%%%%%
\subsection{$\Delta F =2 $ Processes}
%%%%%%%%%%%%%%%%%%%%%%%%

Within the Standard Model, flavor changing neutral currents (FCNC)  are absent at tree-level, and highly suppressed by the GIM mechanism at one loop. Thus, FCNC observables are extremely sensitive to new physics. 
In models of RPV, FCNC operators are generated at tree-level, with the strongest constraints obtained from the $\Delta F =2 $  neutral meson-mixing processes. If either of the operators, $Q_i\bar{u}_j L_k^*$ or  $Q_i Q_j \bar{d}_k^*$, are present,  neutral meson-mixing is generated once the squarks and sleptons are integrated out.   The  diagrams are shown on the right of Fig.~\ref{deltab2} and the corresponding operators generated are,
\begin{eqnarray}
\mathcal{Q}_1^{q_i q_j} &\equiv& -\frac{1}{2\Lambda_{1,ij}^2}  (Q_i^\alpha Q_i^\beta) (Q_j^{\alpha\dagger} Q_j^{\beta\dagger})\,,\\
\mathcal{Q}_4^{q_i q_j} &\equiv& \frac{1}{2\Lambda_{4,ij}^2}  \bar{u}_j^\alpha Q_i^\alpha Q_j^{\beta\dagger} \bar{u}_i^{\beta\dagger}\,.
\end{eqnarray}
Here the suppressions are given by,
\beq
\dfrac{1}{\Lambda_{1,ij}^{2}} = \dfrac{\eta^{\prime\prime}_{iik} \eta^{\prime\prime *}_{jjk}}{m^2_{\tilde{d}_{R,k}}} \epsilon_X^2,~~~~~~~~
\dfrac{1}{\Lambda_{4,ij}^{2}} = \dfrac{|\eta^{\prime}_{ijk}|^2}{m^2_{\tilde{\nu}_{L,k}}}\epsilon_X^2.
\eeq

Taking $m_{\tilde{f}} \simeq$ TeV the bounds from neutral meson mixing are~\cite{UTFit},
\beq \begin{array}{rcl}
\Delta m_K&:& |\eta^{\prime\prime}_{11k} \eta^{\prime\prime *}_{22k} \epsilon_X^2 | \lesssim 10^{-10},  \\
\Delta m_D&:& |\eta^{\prime\prime}_{11k} \eta^{\prime\prime *}_{22k} \epsilon_X^2| \lesssim 10^{-8} , ~~~~ |\eta^{\prime}_{12k} \epsilon_X|^2 \lesssim 10^{-9}\,, \\
\Delta m_{B_d}&:& |\eta^{\prime\prime}_{11k} \eta^{\prime\prime *}_{33k} \epsilon_X^2| \lesssim 10^{-7},  \\
\Delta m_{B_s}&:& |\eta^{\prime\prime}_{23k} \eta^{\prime\prime *}_{33k} \epsilon_X^2| \lesssim 10^{-7}.  \\
\end{array}\eeq
All of the dRPV operators are within the above limits for $\epsilon_X= {\cal O}(10^{-5})$, with or without additional flavor suppressions. 
We note that  the operator $\mathcal{Q}_4^{d_1 d_2}$, which is strongly constrained by $K-\bar{K}$ mixing, is not generated at tree-level in nonholomorphic RPV, while it is in the standard holomorphic case.

\subsection{Proton Decay}
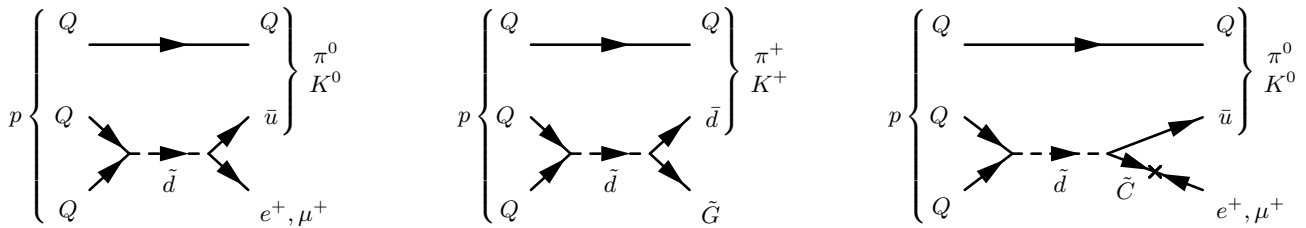
\begin{figure*}
\begin{center}
\begin{fmffile}{pdecay} \hspace*{.5mm}
	        \begin{fmfgraph*}(60,55)
	     \fmfstraight
	        \fmfleft{i1,i2,i3}
	        \fmfright{o1,o2,o3}
                 \fmf{fermion}{i3,o3}
                 \fmf{fermion}{i2,v1}
                 \fmf{fermion}{i1,v1}
                 \fmf{scalar,label=$\tilde{d}$}{v1,v2}
                 \fmf{fermion}{v2,o2}
                 \fmf{fermion}{v2,o1}

                  \fmflabel{$Q$}{i1}
	        \fmflabel{$Q$}{i2}
	        \fmflabel{$Q$}{i3}
	        \fmflabel{$e^+,\mu^+$}{o1}
	        \fmflabel{$\bar{u}$}{o2}
	        \fmflabel{$Q$}{o3}
 
	        \end{fmfgraph*}
	    \end{fmffile}  
\rpar{\begin{array}{c} \pi^0 \\ K^0\end{array}}{10mm}{15mm}{2mm}
\lpar{p}{15mm}{9mm}{-35mm} \hspace*{3cm}
\begin{fmffile}{pdecay2}
	        \begin{fmfgraph*}(60,55)
	     \fmfstraight
	        \fmfleft{i1,i2,i3}
	        \fmfright{o1,o2,o3}
                 \fmf{fermion}{i3,o3}
                 \fmf{fermion}{i2,v1}
                 \fmf{fermion}{i1,v1}
                 \fmf{scalar,label=$\tilde{d}$}{v1,v2}
                 \fmf{fermion}{v2,o2}
                 \fmf{fermion}{v2,o1}

                  \fmflabel{$Q$}{i1}
	        \fmflabel{$Q$}{i2}
	        \fmflabel{$Q$}{i3}
	        \fmflabel{$\tilde{G}$}{o1}
	        \fmflabel{$\bar{d}$}{o2}
	        \fmflabel{$Q$}{o3}
 
	        \end{fmfgraph*}
	    \end{fmffile}  
\rpar{\begin{array}{c} \pi^+ \\ K^+\end{array}}{10mm}{15mm}{2mm}
\lpar{p}{15mm}{9mm}{-34mm}
\hspace*{2cm}
\begin{fmffile}{pdecay3}\hspace*{9mm}
	        \begin{fmfgraph*}(90,55)
	     \fmfstraight
	        \fmfleft{i1,i2,i3}
	        \fmfright{o1,o2,o3}
                 \fmf{fermion}{i3,o3}
                 \fmf{fermion}{i2,v1}
                 \fmf{fermion}{i1,v1}
                 \fmf{scalar,label=$\tilde{d}$}{v1,v2}
                 \fmf{plain}{v2,v3}
                 \fmf{fermion,label=$\tilde{C}$}{v2,v4}	        
                 \fmf{fermion}{v3,o2}
                 \fmf{fermion}{o1,v4}

                  \fmflabel{$Q$}{i1}
	        \fmflabel{$Q$}{i2}
	        \fmflabel{$Q$}{i3}
	        \fmflabel{$e^+,\mu^+$}{o1}
	        \fmflabel{$\bar{u}$}{o2}
	        \fmflabel{$Q$}{o3}
 \fmfv{decoration.shape=cross,decoration.angle=0,decoration.size=3thick}{v4}
	        \end{fmfgraph*}
	    \end{fmffile}  
\rpar{\begin{array}{c} \pi^0 \\ K^0\end{array}}{10mm}{15mm}{2mm}
\lpar{p}{15mm}{9mm}{-45mm}
 
\end{center}
\caption{Diagrams which induce proton decay. {\bf Left:} Decay with L violation via $\eta$ and $\eta'$.  {\bf Center}: Decay to gravitino without L violation.  {\bf Right:} Decay with L violation via the bilinears $\kappa ,\kappa'$.\label{fig:pdecay1}}
\end{figure*}
 
Perhaps the strongest constraint in RPV theories occurs in the case where both B and L are violated (or only B is violated but the gravitino is light),  and as a consequence, the proton becomes unstable. The leading contribution to proton decay comes from the diagram on the left in Fig.~\ref{fig:pdecay1} and gives the decays $p^+ \rightarrow (\pi^0 \rm{~or~} K^0) (e^+ \rm{~or~} \mu^+) $.  The matrix element for the process is,
\beq
\mathcal{M} \simeq  2\eta^*_{m \ell k} \eta^{\prime\prime}_{1 1 k} \epsilon_X^2 \frac{\tilde{\Lambda}^2_{QCD}}{m^2_{\tilde{d}_{R,k}}}\,,
\eeq
where $m=1 (2)$ for a pion(kaon) and $\ell=1 (2)$ for electron(muon). Taking, as before, $\tilde{\Lambda}_{QCD} = 200$ MeV, one finds a lifetime of order, 
\beq
\tau_p \simeq 5 \times10^{33} {\rm yr}   \left(\frac{m_{\tilde{d}_{R\,k}}}{{\rm TeV}} \right)^4
 \left(\frac{10^{-14}}{|\eta_{m \ell k} \eta^{\prime\prime}_{1 1 k}|} \right)^2 \left( \frac{10^{-5}}{\epsilon_X}\right)^4\ .
\eeq 
The above should be compared to the relevant limit.  The strongest is found for the $p\to e^+\pi^0$ decay mode, $\tau_p > 8.2\times 10^{33}$ years~\cite{Beringer:1900zz}.  

Another channel for proton decay can appear if the gravitino is light, leading to the decays $p\to (\pi ,K)+\tilde{G}$ as shown in the center of  Fig.~\ref{fig:pdecay1}. The matrix element is estimated to be 
\begin{equation}
|{\cal M}|^2 \sim \frac{1}{3} |\eta_{11i}^{\prime\prime}|^2 \epsilon_X^2 \frac{m_p^4 \tilde{\Lambda}_{QCD}^4}{m_{\tilde{d}_i}^4 m_{3/2}^2 M_{Pl}^2}\,,
\end{equation}
and the corresponding lifetime is
\begin{equation}
\tau_p \sim 2 \times 10^{33} {\rm yr} \left( \frac{m_{\tilde{d}_i}}{\rm TeV}\right)^4 \left( \frac{M}{ {10^8 \rm GeV}}\right)^4 \left( \frac{10^{-8}}{|\eta_{11i}^{\prime\prime}|}\right)^2 \left(\frac{F}{F_X} \right)^2\,.
\end{equation} 
Here $F=  \sqrt{3} m_{3/2} M_{\rm pl}$, while $F_X$ denotes, as above, the F-term for $X$.   In the case of a single SUSY-breaking sector one has $F=F_X$; more generally, additional sources of SUSY-breaking can exist and will relax the constraint.  In this case, the strongest constraint is obtained from a search for $p\to \nu K$ which gives~\cite{Beringer:1900zz}, $\tau_p > 2.3\times 10^{33}$ years.

Finally, proton decay may also result from the lepton number violating operators parameterized by $\kappa , \kappa'$. These will induce mass and kinetic mixing between the charged leptons and the charginos (via their charged  higgsino components). The mixing will lead to proton decay through the diagram shown on the right of Fig.~\ref{fig:pdecay1}. The decay amplitude is given by,
\beq
\mathcal{M} \simeq   \eta^{\prime\prime}_{1 1 k} \kappa^{\rm eff}_k \epsilon_X \frac{\tilde{\Lambda}^2_{QCD}}{m^2_{\tilde{d}_{R,k}}}\,,
\eeq
where $\kappa^{\rm eff}_k = \kappa_k \frac{v_d}{M} + \kappa_k  \epsilon_X \frac{m_{e_k} v_u}{m_{\tilde{C}}} +\kappa'_k \frac{M}{M_{Pl}}$ defines the effective mixing between the electron and the chargino. The resulting proton lifetime is
\beq
\tau_p \simeq 3\times10^{33} {\rm yr}   \left(\frac{m_{\tilde{d}_{R\,k}}}{{\rm TeV}} \right)^4
 \left(\frac{3\times 10^{-20}}{|\kappa^{\rm eff}_{k}\eta^{\prime\prime}_{1 1 k}|} \right)^2 \left( \frac{10^{-5}}{\epsilon_X}\right)^2\ .
\eeq

\subsection{Cosmology}

Rapid $B$ and $L$ violating interactions induced by RPV operators may wash out any pre-existing baryonic or leptonic asymmetry.   Consequently, such processes should be highly suppressed at low temperatures.   
Since 
sphalerons, active above the weak-scale, violate $B+L$,
 it is typically required that the RPV-induced rates are sufficiently slow above that scale.   The bounds on the dRPV operators are similar to those in  
standard holomorphic RPV.   One finds $\epsilon_X \eta \lesssim 10^{-7}$ and $\kappa_i^{\rm eff}<10^{-6}$ where $\eta$ stands for any $\eta_{ijk}$, $\eta'_{ijk}$ or $\eta''_{ijk}$~\cite{RPVreview,Campbell:1991at,Endo:2009cv}.

As we show below, these cosmological bounds  typically imply  displaced decays at the LHC.  Nonetheless these bounds can be easily evaded in several ways (see~\cite{RPVreview} and references therein).   For example,  the bounds are irrelevant if the baryon asymmetry is generated at or below the electroweak scale.  Conversely,  as discussed in~\cite{Endo:2009cv,Ruderman:2012jd}, when a single lepton flavor number is approximately conserved the bounds can be significantly weaker.

%%%%%%%%%%%%%%%%%%%%%%%%%%%%%%%%%%%%%%
\section{LHC phenomenology}

The phenomenology of  models with dRPV can be very different from those with R-parity conservation and even from those with traditional RPV described by (\ref{eq:WRPV}).   The details
depend greatly on the identity of the lightest supersymmetric particle (LSP). Here we briefly comment on three interesting possibilities which crucially differ in their collider phenomenology from standard RPV: stop LSP, gluino LSP and sneutrino LSP, with the first two most relevant for naturalness.   
Further details on these and other interesting possibilities will be given in~\cite{future1}. 

\begin{figure}
\begin{center}
\includegraphics[width=0.4\textwidth]{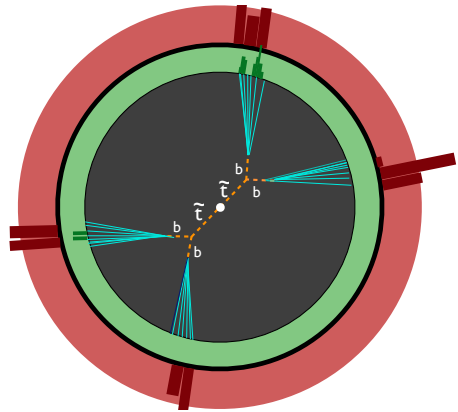}
\end{center}
\caption{An illustrative event display for stop pair production, decaying via dRPV operators to 4b.    Both the bottoms and the stops decay a finite distance from their production vertex.   Each of the bottom pairs reconstructs to a single displaced vertex with a stop invariant mass.
\label{fig:stopLSP}}
\end{figure}

Consider first the stop LSP.   In all of the non-holomorphic operators of (\ref{eq:KRPV}),  stop decays are induced from SUSY-conserving interactions in which the stop is extracted from one of the chiral fields.  As a consequence, the resulting operators in the Lagrangian all have derivative couplings and hence the decay rate is chirally suppressed.     One finds that the dominant decay mode is typically $\tilde t\to \bar b \bar b$,  with a decay length,
\begin{equation}
c \tau_{\tilde{t}} \simeq 1 {\rm~mm} \left(\frac{300  {\rm ~GeV}}{{m}_{\tilde{t}}} \right) \left(\frac{M}{ 10^8{\rm  GeV}} \right)^2
\left|\frac{1}{\eta_{333}^{\prime\prime}} \right|^2 \,.
\end{equation}
Thus the stop LSP case may manifest itself uniquely as four displaced b's, where each pair reconstructs to a single displaced vertex, and the two pairs have a similar invariant mass.  The situation is illustrated in Fig.~\ref{fig:stopLSP}.   We stress that such decays do not exist in the holomorphic RPV scenario.     {\em The collider search for a stop LSP should be significantly altered in order to discover dRPV.}

Next consider the case of a sneutrino LSP, where the LSP decay is governed by the $\eta'$ couplings which induce the operators $u_{Li} u_{Rj}^\dagger \tilde{\nu}_k + d_{Li} u_{Rj}^\dagger \tilde{e}_{Lk}^\dagger$. Since the 3rd generation couplings are typically least suppressed, the leading decay mode will be $\tilde{\nu}\to t_L t_R^\dagger$ with a decay length
\begin{equation}
c \tau_{\tilde \nu} \simeq 1\ {\rm mm} \ \left|\frac{10^{-2} }{\eta'_{331}}\right|^2 \left( \frac{10^{-5}}{\epsilon_X}\right)^2 \frac{ 165 \ {\rm GeV}}{(1-2\frac{m_t^2}{m_{\tilde{\nu}}^2})\sqrt{m_{\tilde{\nu}}^2-4{m_t^2}}}.
\end{equation} 
For $\eta'_{331} \lesssim 10^{-2}$ this vertex will be displaced, leading to the interesting LHC signal of 4 displaced top quarks in the final state. 

Finally, a gluino LSP decays via an off-shell stop to two bottoms and a top, $\tilde g \to t  b  b$.  The decay length here is estimated at,
\begin{equation}
\label{eq:gluinoLSP}
c \tau_{\tilde{g}} \simeq 1~{\rm mm} \left|\frac{1 }{\eta''_{333}}\right|^2 \left( \frac{m_{\tilde{t}}}{400 {\rm GeV}} \right)^4
 \left( \frac{350{\rm GeV}}{m_{\tilde{g}}} \right)^5   \left( \frac{M}{10^6{\rm GeV}}\right)^2\,.
\end{equation}
A late decaying gluino is less constrained than a promptly decaying one. This possibility may allow for a lighter gluino to be produced at the LHC~\cite{future1}.
%%%%%%%%%%%%%%%%%%%%%%%%%%%%%%%%%%%%%%

%\begin{acknowledgments}
{\it\bf Acknowledgments:} 
We thank Yonit Hochberg and Abner Soffer for useful discussions.   We especially thank Ben Heidenreich for many useful comments on the manuscript.
C.C. is supported in part by the NSF grant PHY-0757868. 
E.K. and T.V. are supported in part by a grant from the Israel Science Foundation.  T.V. is further supported by the US-Israel Binational Science Foundation,  the EU-FP7 Marie Curie, CIG fellowship and  by the I-CORE Program of the Planning and Budgeting Committee and The Israel Science Foundation (grant NO 1937/12).
 C.C. thanks the particle theory group at Tel Aviv University, the Galileo Galilei Institute for Theoretical Physics/INFN and the Aspen Center for Physics for their hospitality while part of this work was completed. 
%\end{acknowledgements}
%%%%%%%%%%%%%%%%%%%%%%%%%%%%%%%%%%%%%%%%%%%%%%

 %%%%%%%%%%%%%%%%%%%%%%%%%%%%%%%%%%%%%%%%%%%%%%
%%%%%%%%%%%%%%%%%%%%%%%%%%%%%%%%%%%%%%%%%%%%%%

\begin{thebibliography}{0}

  \bibitem{RPV}
L.~J.~Hall and M.~Suzuki,
  %``Explicit R-Parity Breaking in Supersymmetric Models,''
  Nucl.\ Phys.\ B {\bf 231}, 419 (1984);
  %%CITATION = NUPHA,B231,419;%%
 G.~G.~Ross and J.~W.~F.~Valle,
  %``Supersymmetric Models Without R-Parity,''
  Phys.\ Lett.\ B {\bf 151}, 375 (1985);
  %%CITATION = PHLTA,B151,375;%%
 V.~D.~Barger, G.~F.~Giudice and T.~Han,
  %``Some New Aspects of Supersymmetry R-Parity Violating Interactions,''
  Phys.\ Rev.\ D {\bf 40}, 2987 (1989);
  %%CITATION = PHRVA,D40,2987;%%
H.~K.~Dreiner,
  %``An Introduction to explicit R-parity violation,''
  In *Kane, G.L. (ed.): Perspectives on supersymmetry II* 565-583
  [hep-ph/9707435];
  %%CITATION = HEP-PH/9707435;%%
 G.~Bhattacharyya,
  %``A Brief review of R-parity violating couplings,''
  In *Tegernsee 1997, Beyond the desert 1997* 194-201
  [hep-ph/9709395].
  %%CITATION = HEP-PH/9709395;%%

\bibitem{RPVreview}
 R.~Barbier, C.~Berat, M.~Besancon, M.~Chemtob, A.~Deandrea, E.~Dudas, P.~Fayet and S.~Lavignac {\it et al.},
  %``R-parity violating supersymmetry,''
  Phys.\ Rept.\  {\bf 420}, 1 (2005)
  [hep-ph/0406039].
  %%CITATION = HEP-PH/0406039;%%

\bibitem{MFVSUSY}
C.~Csaki, Y.~Grossman and B.~Heidenreich,
  %``MFV SUSY: A Natural Theory for R-Parity Violation,''
  Phys.\ Rev.\ D {\bf 85}, 095009 (2012)
  [arXiv:1111.1239 [hep-ph]].
  %%CITATION = ARXIV:1111.1239;%%

\bibitem{Smith}
 E.~Nikolidakis and C.~Smith,
  %``Minimal Flavor Violation, Seesaw, and R-parity,''
  Phys.\ Rev.\ D {\bf 77}, 015021 (2008)
  [arXiv:0710.3129 [hep-ph]].
  %%CITATION = ARXIV:0710.3129;%%

\bibitem{Dreiner}
  H.~K.~Dreiner and M.~Thormeier,
  %``Supersymmetric Froggatt-Nielsen models with baryon and lepton number violation,''
  Phys.\ Rev.\ D {\bf 69}, 053002 (2004)
  [hep-ph/0305270].
  %%CITATION = HEP-PH/0305270;%%
  %47 citations counted in INSPIRE as of 13 Sep 2013
  A.~Monteux,
  %``Natural, R-parity violating supersymmetry and horizontal flavor symmetries,''
  Phys.\  Rev.\  D 88, {\bf 045029} (2013)
  [Phys.\ Rev.\ D {\bf 88}, 045029 (2013)]
  [arXiv:1305.2921 [hep-ph]].
  %%CITATION = ARXIV:1305.2921;%%
  %1 citations counted in INSPIRE as of 14 Sep 2013


\bibitem{moreRPVflavor} 
  B.~Keren-Zur, P.~Lodone, M.~Nardecchia, D.~Pappadopulo, R.~Rattazzi and L.~Vecchi,
  %``On Partial Compositeness and the CP asymmetry in charm decays,''
  Nucl.\ Phys.\ B {\bf 867}, 429 (2013)
  [arXiv:1205.5803 [hep-ph]];
  %%CITATION = ARXIV:1205.5803;%%
  %30 citations counted in INSPIRE as of 19 Jun 2013
  G.~Krnjaic and D.~Stolarski,
  %``Gauging the Way to MFV,''
  arXiv:1212.4860 [hep-ph];
  %%CITATION = ARXIV:1212.4860;%%
 R.~Franceschini and R.~N.~Mohapatra,
  %``New Patterns of Natural R-Parity Violation with Supersymmetric Gauged Flavor,''
  arXiv:1301.3637 [hep-ph];
  %%CITATION = ARXIV:1301.3637;%%
  C.~Csaki and B.~Heidenreich,
  %``A Complete Model for R-parity Violation,''
  arXiv:1302.0004 [hep-ph];
  %%CITATION = ARXIV:1302.0004;%%
  %4 citations counted in INSPIRE as of 19 Jun 2013
  
  
  \bibitem{Krnjaic:2013eta}
  G.~Krnjaic and Y.~Tsai,
  %``Soft RPV Through the Baryon Portal,''
  arXiv:1304.7004 [hep-ph].
  %%CITATION = ARXIV:1304.7004;%%
  %1 citations counted in INSPIRE as of 19 Jun 2013
  %\cite{Acharya:2011te}
%\bibitem{Acharya:2011te} 
  B.~S.~Acharya, G.~Kane, E.~Kuflik and R.~Lu,
  %``Theory and Phenomenology of $\mu$ in M theory,''
  JHEP {\bf 1105}, 033 (2011)
  [arXiv:1102.0556 [hep-ph]].
  %%CITATION = ARXIV:1102.0556;%%
  %10 citations counted in INSPIRE as of 22 Sep 2013

\bibitem{RPVLHC}
 C.~Brust, A.~Katz and R.~Sundrum,
  %``SUSY Stops at a Bump,''
  JHEP {\bf 1208}, 059 (2012)
  [arXiv:1206.2353 [hep-ph]];
  %%CITATION = ARXIV:1206.2353;%%
 P.~W.~Graham, D.~E.~Kaplan, S.~Rajendran and P.~Saraswat,
  %``Displaced Supersymmetry,''
  JHEP {\bf 1207}, 149 (2012)
  [arXiv:1204.6038 [hep-ph]].
  %%CITATION = ARXIV:1204.6038;%%
 P.~Fileviez Perez and S.~Spinner,
  %``The Minimal Theory for R-parity Violation at the LHC,''
  JHEP {\bf 1204}, 118 (2012)
  [arXiv:1201.5923 [hep-ph]];
  %%CITATION = ARXIV:1201.5923;%%
J.~A.~Evans and Y.~Kats,
  %``LHC Coverage of RPV MSSM with Light Stops,''
  arXiv:1209.0764 [hep-ph];
  %%CITATION = ARXIV:1209.0764;%%
Z.~Han, A.~Katz, M.~Son and B.~Tweedie,
  %``Boosting Searches for Natural SUSY with RPV via Gluino Cascades,''
  arXiv:1211.4025 [hep-ph];
  %%CITATION = ARXIV:1211.4025;%%
 J.~Berger, C.~Csaki, Y.~Grossman and B.~Heidenreich,
  %``Mesino Oscillation in MFV SUSY,''
  arXiv:1209.4645 [hep-ph];
  %%CITATION = ARXIV:1209.4645;%%
  R.~Franceschini and R.~Torre,
  %``RPV stops bump off the background,''
  arXiv:1212.3622 [hep-ph].
  %%CITATION = ARXIV:1212.3622;%%

\bibitem{Ruderman:2012jd}
J.~T.~Ruderman, T.~R.~Slatyer and N.~Weiner,
  %``A Collective Breaking of R-Parity,''
  arXiv:1207.5787 [hep-ph];
  %%CITATION = ARXIV:1207.5787;%%
\bibitem{FN}
 C.~D.~Froggatt and H.~B.~Nielsen,
  %``Hierarchy of Quark Masses, Cabibbo Angles and CP Violation,''
  Nucl.\ Phys.\ B {\bf 147}, 277 (1979).
  %%CITATION = NUPHA,B147,277;%%
  %1199 citations counted in INSPIRE as of 09 Aug 2013
 \bibitem{flavormediation} 

D.~E.~Kaplan, F.~Lepeintre, A.~Masiero, A.~E.~Nelson and A.~Riotto,
  %``Fermion masses and gauge mediated supersymmetry breaking from a single U(1),''
  Phys.\ Rev.\ D {\bf 60}, 055003 (1999)
  [hep-ph/9806430];
  %%CITATION = HEP-PH/9806430;%%
  %45 citations counted in INSPIRE as of 17 Sep 2013
  D.~E.~Kaplan and G.~D.~Kribs,
  %``Phenomenology of flavor mediated supersymmetry breaking,''
  Phys.\ Rev.\ D {\bf 61}, 075011 (2000)
  [hep-ph/9906341].
  %%CITATION = HEP-PH/9906341;%%
  %24 citations counted in INSPIRE as of 13 Sep 2013

\bibitem{future1}
C.~Csaki, E.~Kuflik and T.~Volansky, to appear. 




\bibitem{RS}
 L.~Randall and R.~Sundrum,
  %``A Large mass hierarchy from a small extra dimension,''
  Phys.\ Rev.\ Lett.\  {\bf 83}, 3370 (1999)
  [hep-ph/9905221];
  %%CITATION = HEP-PH/9905221;%%
  %5869 citations counted in INSPIRE as of 09 Aug 2013
 S.~J.~Huber and Q.~Shafi,
  %``Fermion masses, mixings and proton decay in a Randall-Sundrum model,''
  Phys.\ Lett.\ B {\bf 498}, 256 (2001)
  [hep-ph/0010195].
  %%CITATION = HEP-PH/0010195;%%
  %305 citations counted in INSPIRE as of 09 Aug 2013
K.~Agashe, G.~Perez and A.~Soni,
  %``Flavor structure of warped extra dimension models,''
  Phys.\ Rev.\ D {\bf 71}, 016002 (2005)
  [hep-ph/0408134];
  %%CITATION = HEP-PH/0408134;%%
  %300 citations counted in INSPIRE as of 09 Aug 2013
C.~Csaki, A.~Falkowski and A.~Weiler,
  %``The Flavor of the Composite Pseudo-Goldstone Higgs,''
  JHEP {\bf 0809}, 008 (2008)
  [arXiv:0804.1954 [hep-ph]].
  %%CITATION = ARXIV:0804.1954;%%
  %161 citations counted in INSPIRE as of 09 Aug 2013
/9310320;%%<br /> 372 citations counted in INSPIRE as of 02 Sep 2013



\bibitem{NelsonStrassler}
A.~E.~Nelson and M.~J.~Strassler,
  %``Suppressing flavor anarchy,''
  JHEP {\bf 0009}, 030 (2000)
  [hep-ph/0006251].
  %%CITATION = HEP-PH/0006251;%%
  %118 citations counted in INSPIRE as of 09 Aug 2013


\bibitem{Rattazzi}
 R.~Rattazzi and A.~Zaffaroni,
  %``Comments on the holographic picture of the Randall-Sundrum model,''
  JHEP {\bf 0104}, 021 (2001)
  [hep-th/0012248].
  %%CITATION = HEP-TH/0012248;%%
  %306 citations counted in INSPIRE as of 09 Aug 2013


\bibitem{NirSeiberg}
M.~Leurer, Y.~Nir and N.~Seiberg,
``Mass Matrix Models: the Sequel,''
Nucl.\ Phys.\ B {\bf 420} (1994) 468
[hep-ph/9310320].
%%CITATION = HEP-PH


%\cite{Abe:2011ky}
\bibitem{Abe:2011ky} 
  K.~Abe {\it et al.}  [Super-Kamiokande Collaboration],
  %``The Search for $n-\bar{n}$ oscillation in Super-Kamiokande I,''
  arXiv:1109.4227 [hep-ex].
  %%CITATION = ARXIV:1109.4227;%%
  %8 citations counted in INSPIR



\bibitem{Sher}
J.~L.~Goity and M.~Sher,
  %``Bounds on delta B = 1 couplings in the supersymmetric standard model,''
  Phys.\ Lett.\ B {\bf 346}, 69 (1995)
  [Erratum-ibid.\ B {\bf 385}, 500 (1996)]
  [hep-ph/9412208].
  %%CITATION = HEP-PH/9412208;%%
  %147 citations counted in INSPIRE as of 25 Jun 2013
E as of 02 Jul 2013
  
\bibitem{Litos:2010zra} 
  M.~D.~Litos, ``A search for dinucleon decay into kaons using the  Super-Kamiokande water Cherenkov detector,''
  %%CITATION = INSPIRE-1224703;%%

\bibitem{UTFit}
 M.~Bona {\it et al.}  [UTfit Collaboration],
  %``Model-independent constraints on $\Delta$ F=2 operators and the scale of new physics,''
  JHEP {\bf 0803}, 049 (2008)
  [arXiv:0707.0636 [hep-ph]].
  %%CITATION = ARXIV:0707.0636;%%
  %271 citations counted in INSPIRE as of 09 Aug 2013



\bibitem{Beringer:1900zz}
J.~Beringer {\it et al.} [Particle Data Group Collaboration],
``Review of Particle Physics (Rpp),''
Phys.\ Rev.\ D {\bf 86} (2012) 010001.
%%CITATION = PHRVA,D86,010001;%%<br /> 2113 citations counted in INSPIRE as of 02 Sep 2013
  %\cite{Litos:2010zra}

\bibitem{Campbell:1991at} 
  B.~A.~Campbell, S.~Davidson, J.~R.~Ellis and K.~A.~Olive,
  %``On B+L violation in the laboratory in the light of cosmological and astrophysical constraints,''
  Astropart.\ Phys.\  {\bf 1}, 77 (1992).
  %%CITATION = APHYE,1,77;%%
  %42 citations counted in INSPIRE as of 17 Sep 2013 
  
  \bibitem{Endo:2009cv}
M.~Endo, K.~Hamaguchi and S.~Iwamoto,
``Lepton Flavor Violation and Cosmological Constraints on R-Parity Violation,''
JCAP {\bf 1002} (2010) 032
[arXiv:0912.0585 [hep-ph]].
%%CITATION = ARXIV:0912.0585;%%<br /> 15 citations counted in INSPIRE as of 02 Sep 2013
   
 \end{thebibliography}
\end{document}